\documentclass[]{aa}  

\usepackage{graphicx}
\usepackage{subfig}
\usepackage{txfonts}
\usepackage[utf8]{inputenc}
\usepackage[bookmarks=false, colorlinks=true, citecolor=blue, linkcolor=blue]{hyperref}
\def\kms{\,km\,s$^{-1}$}

\begin{document}

\title{Detection of periodic flares in 6.7\,GHz methanol masers G45.804$-$0.356 and G49.043$-$1.079 \thanks{The data used to produce Fig. 3 are available in electronic form at the CDS via anonymous ftp to cdsarc.u-strasbg.fr (130.79.128.5) or via http://cdsweb.u-strasbg.fr/cgi-bin/qcat?J/A+A/}}
\titlerunning{Periodicity of 6.7\,GHz line in G45.804$-$0.356 and G49.043$-$1.079}
\authorrunning{M. Olech et al.}

\author{M. Olech 
          \inst{1} \href{https://orcid.org/0000-0002-0324-7661}{\includegraphics[scale=0.5]{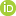}}
          \and
          M. Durjasz\inst{2} \href{https://orcid.org/0000-0001-7952-0305}{\includegraphics[scale=0.5]{orcid.png}}%\fnmsep
          \and
          M. Szymczak\inst{2} \href{https://orcid.org/0000-0002-1482-8189}{\includegraphics[scale=0.5]{orcid.png}}%\fnmsep
          \and
                  A. Bartkiewicz \inst{2} \href{https://orcid.org/0000-0002-6466-117X}{\includegraphics[scale=0.5]{orcid.png}}
}

    \institute{Space Radio-Diagnostic Research Center, Faculty of Geoengineering, University of Warmia and Mazury\\  ul.Oczapowskiego 2, PL-10-719 Olsztyn, Poland
        \and Institute of Astronomy, Faculty of Physics, Astronomy and Informatics, Nicolaus Copernicus University,\\
   Grudziadzka 5, 87-100 Torun, Poland
    }

   \date{Received month day , year; accepted month day, year}

\abstract
{Periodicity in 6.7\,GHz methanol maser sources is a rare phenomenon that was discovered during long-term monitoring programmes. Understanding  the underlying processes that lead to periodic variability might provide insights into the physical processes in high-mass star-forming regions.}  
{We aim to identify and describe  new periodic methanol masers.} 
{The observations were obtained with the Torun 32m antenna. Time series analysis was conducted using well-proven statistical methods. Additionally, NEOWISE data were used to search for a correlation between infrared and maser fluxes.} 
{We found two new periodic sources, G45.804$-$0.356 and G49.043$-$1.079, with periods of 416.9 and 469.3 days, respectively. For  G49.043$-$1.079, infrared variability is simultaneous with methanol flares.}
{A most likely cause of the periodicity in G49.043$-$1.079 is modulated accretion. For G45.804$-$0.356, the periodicity cannot be explained with the available data, and further research is needed.}

  \keywords{masers -- stars:formation -- ISM:clouds -- radio lines:ISM --  individual: G45.804$-$0.356, G49.043$-$1.079}

   \maketitle
%
%-------------------------------------------------------------------
% This text is not a text of honor... no highly esteemed deed is commemorated here... nothing valued is here.
% What is here was dangerous and repulsive to us. This message is a warning about danger. 

\section{Introduction}
Emission of class II 6.7\,GHz methanol maser lines is a well-known tracer of high-mass star-forming regions \citep{Menten1991,Breen2015,ellingsen2006,Pandian2007,Szymczak2012}. It is radiatively pumped by mid-infrared photons emitted by dust grains in the local environment \citep{Cragg2005,Urquhart2015}. 
Milliarcsecond imaging of 6.7\,GHz sources shows that emission arises in most cases from regions of an interface between the protostellar disc and envelope at a typical distance of $\sim$1000~au from a protostar with individual maser cloudlets  of size $\sim$10 au  \citep{Goddi2011,bartkiewicz2016}. The 6.7\,GHz line is sensitive to the local physical conditions, and its location makes it a perfect tool for studying processes during this early stage of high-mass star formation. Recent studies of strong methanol maser flares have shown that observations of this line can be used to detect early stages of accretion bursts in massive protostars \citep{Moscadelli2017,Hunter2018,burns2020}.

Several long-term monitoring programmes have been conducted since the discovery of this transition \citep{Goedhart2004,Szymczak2018}. One of their most surprising results was the identification of a small group of 26 sources that show periodic behaviour (e.g. \citealt{Goedhart2003,Maswanganye2015,Szymczak2011,Sugiyama2017}). The periods range from 24 to 670 days, with most sources having periods between 100 and 300 days. The variety in flare profiles and periods has led to a number of competing hypotheses on the nature of this process. Some models, such as the colliding wind binary (CWB) model, consider changes in the seed photon flux to be the primary driving mechanism \citep{Van2009,vanderWalt2011}. Others consider the change in the pumping efficiency to be the main mechanism responsible for flares, such as pulsating protostars due to high accretion \citep{Inayoshi2013}, {a} binary system with spiral shocks \citep{Parfenov2014}, or modulated accretion in binaries \citep{Araya2010}. Although these models were successfully applied to some known sources, there is no clear consensus on the general mechanism of periodicity. Furthermore, \cite{morgan2021} have shown that the orientation of the disc-outflow system in the plane of the sky might have a significant impact on observed flare profiles. The recent discovery of anti-correlated flares of 6.7\,GHz methanol and 22\,GHz water vapour maser lines in  G107.298$+$5.639 and the synchronicity of infrared (IR) and methanol flares in two sources \citep{Szymczak2015,olech2020}  is strong evidence for pumping mechanism modulation being a leading cause. 

Nevertheless, testing current theories is limited by the small number of sources. Extending the list of periodic masers is crucial for facilitating this work. 
Therefore, monitoring programmes of methanol masers are valuable as they can lead to the identification of new periodic sources  \citep{Sugiyama2018_IAU,Szymczak2018,Olech2019}. 

In this paper we report the discovery of periodicity in another two 6.7\,GHz methanol maser sources: G45.804$-$0.356 and G49.043$-$1.079\footnote{The names are the Galactic coordinates.}. 

  \begin{figure}
   \centering
   \includegraphics[width=\columnwidth]{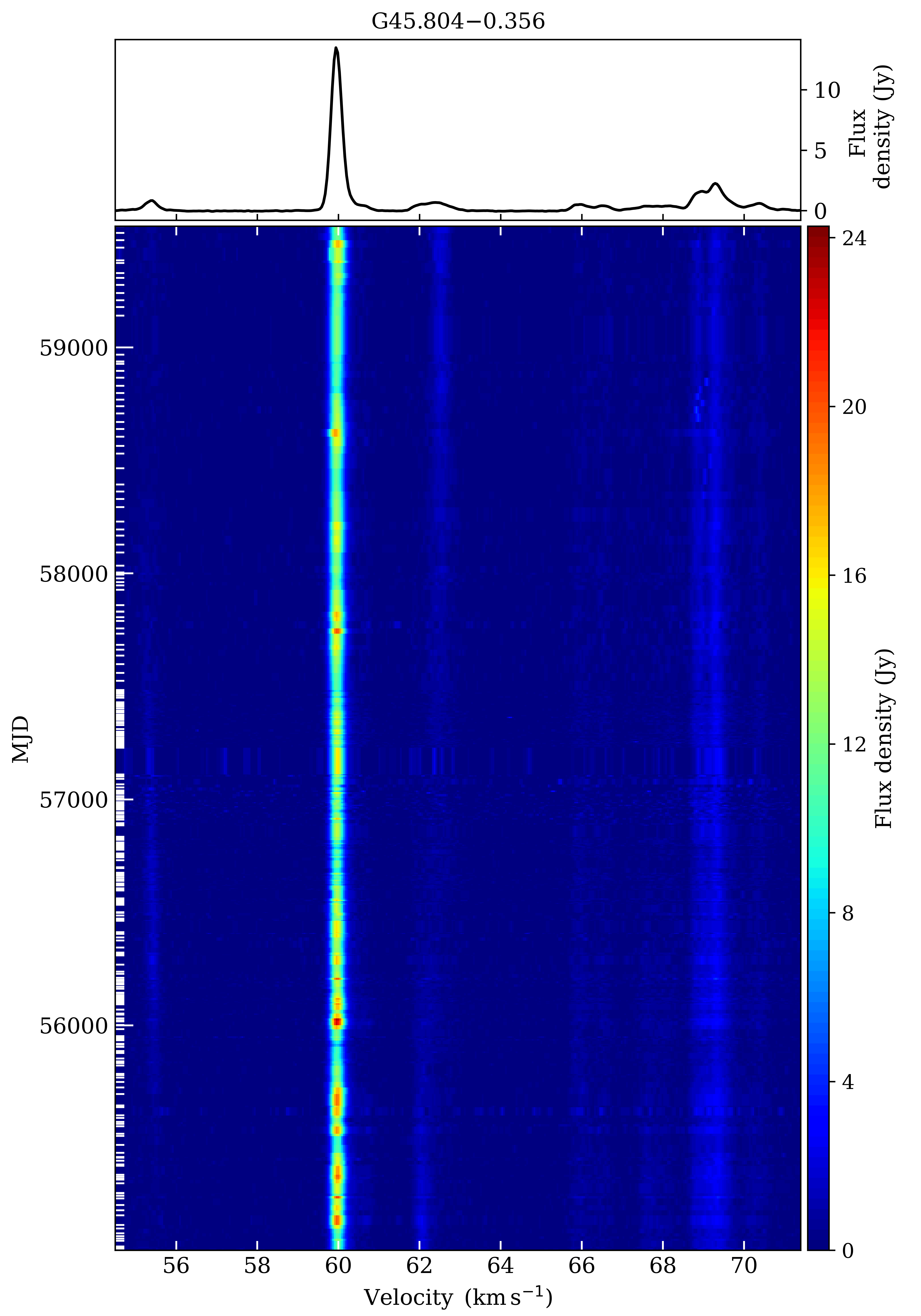}
   \caption{Dynamic spectrum of  G45.804$-$0.356. The velocity scale is relative to the local standard of rest. The horizontal bars in the left coordinate correspond to the dates of the observed spectra. The top panel shows the average spectrum from the whole observation period.}
  \label{fig:g45_dynamic}
 \end{figure}

 \begin{figure}
   \centering
   \includegraphics[width=\columnwidth]{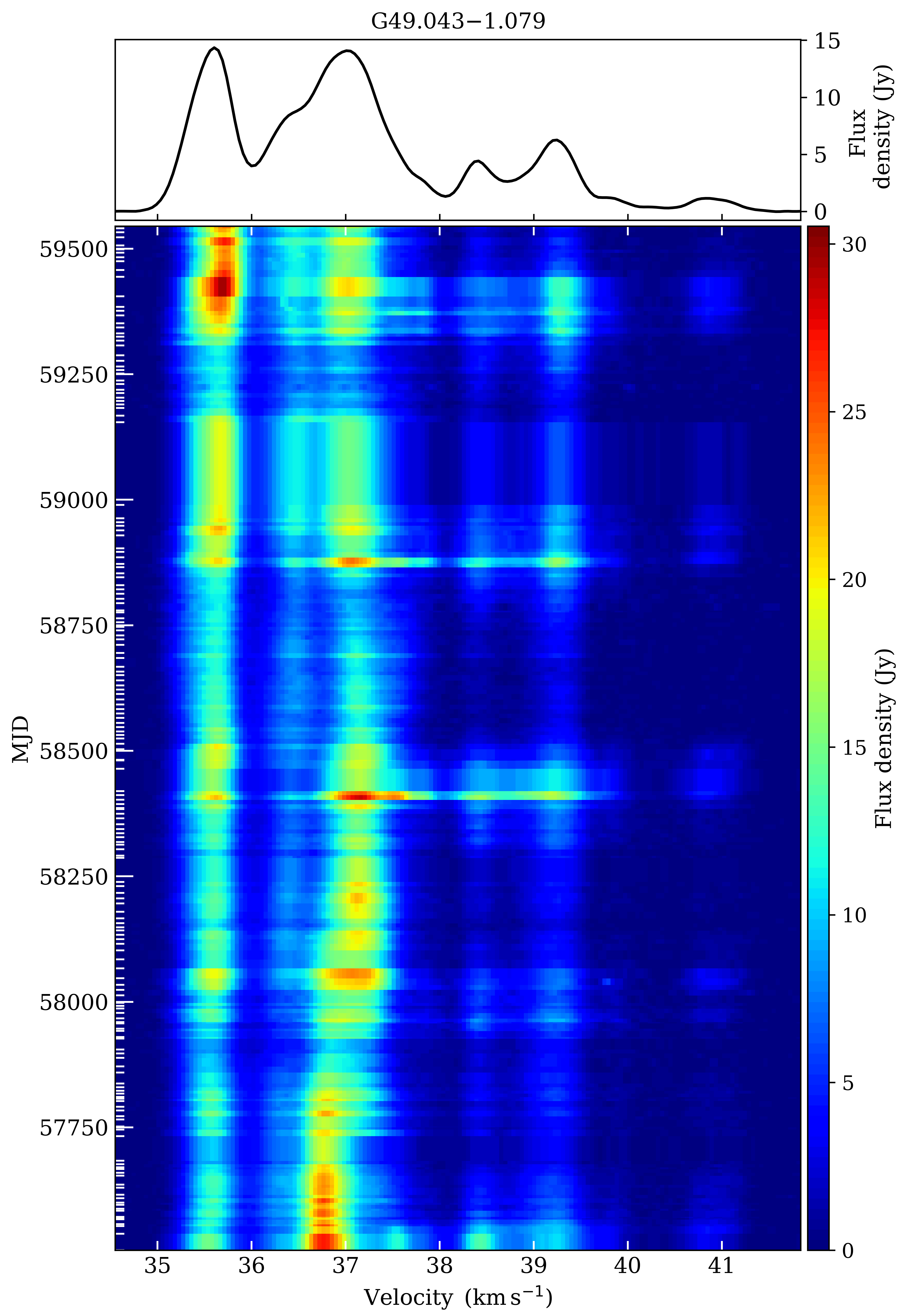}
   
   \caption{Same as Fig.\ref{fig:g45_dynamic} but for  G49.043$-$1.079.}
   \label{fig:g49_dynamic}
 \end{figure}

\section{Observations and methodology}
%\subsection{Observations}
Both targets were observed with the Torun 32m antenna as part of an ongoing long-term monitoring programme of methanol maser sources \citep{Szymczak2018}. G45.804$-$0.356 
(RA=19$^{\mathrm{h}}$16$^{\mathrm{m}}$31$\fs$08, Dec=11$^{o}$16$'$12\farcs0 (J2000)) was observed from June 2009 to October 2021, with a cadence of 1.8 observations per month. For G49.043$-$1.079  (RA=19$^\mathrm{h}$25$^{\mathrm{m}}$22$\fs$20, Dec=13$^{o}$47$'$19\farcs5 (J2000)), monitoring started in April 2016 and ended in November 2021, with a cadence of 2.4 observations per month. Coordinates used for the observations were taken from \cite{Breen2015}.

Scans at the rest frequency of 6668.519~MHz were obtained with the frequency-switching mode. The auto-correlator banks were configured with 4196 channels for the 4\,MHz band per circular polarization. This yielded a spectral resolution of 0.09\kms\, after Hanning smoothing. The system equivalent flux density was measured by observing the strong continuum source 3C123. Furthermore, daily sensitivity variations were calibrated by observing the maser source G32.744$-$0.076, which has known stable features. This scheme had been successfully used in previous studies. We estimate the error of flux calibrations to be 10$\%$. \\

Before May 2015, the typical system temperature ($T_\mathrm{sys}$) was 50\,K with 3$\sigma_{\mathrm{rms}}$ of $\sim$ 1.2 Jy for observations consisting of 45 scans of 30\,s duration each. In May 2015, the new 6.7\,GHz receiver was installed on the antenna, and the average system temperature -- lowered to 30\,K with a 3$\sigma_{\mathrm{rms}}$ of $\sim$ 0.9\,Jy -- could be achieved after only 30 scans. During data reduction, observations with $T_\mathrm{sys}$ much higher than typical as well as observations contaminated with Radio Frequency Interference (RFI) or instrumental effects were rejected.\\

In this work we also used the NEOWISE 2021 Data Release \citep{Mainzer2011,Mainzer2014,Cutri2015}. We used a cone search with a radius of 5" around source positions. To ensure high quality, we discarded all measurements with  $ph\_qual$ other than `A', $cc\_flags$ other than 0, and all $qual\_frame$=0. In total, 190 measurements for G45.804$-$0.356 and 176 for G49.043$-$1.079 were used for the analysis. 

\subsection*{Time series analysis}
A time series of individual spectral features was extracted as the velocity integrated flux.  
Before analysis, all non-periodic trends were removed by fitting the polynomial to the lower envelope of the time series using the least squares method and then subtracting it.
The periodicity was estimated by calculating the Lomb-Scargle (L-S) periodogram \citep{Scargle} using the Astropy Python library \citep{astropy:2013, astropy:2018} and with a simple analysis of variance (AoV) implementation \citep{aov}. The period value was calculated by fitting the Gaussian function to the L-S and AoV maxima, and the error was estimated as its full width at half maximum (FWHM). With the acquired period, the time series was then phased and binned. An asymmetric periodic power function given by the equation 
\begin{center}
 $S(t)= D*\mathrm{exp}(s(t))+C $,\\
 
 where
 $s(t)= -b\,\mathrm{cos}(\omega t+\phi)/(1-f\mathrm{sin}(\omega t+\phi))+a$,
\end{center}

%where 
\noindent
 was fitted to these data. Here, a, b, C, and D are constants and $\omega=2\pi/P$, where $P$ is the period, $\phi$ is the phase, and $f$ is the asymmetry parameter \citep{Szymczak2011}. For phased data, we assumed $P$=1.\\

Average characteristics of the flare, such as the relative amplitude, the flare FWHM, and the rise-to-decay ratio, were estimated from this fitted function. The time lag between the features was calculated using the discrete correlation function \citep[DCF;][]{edelson1988}. Lag estimation was done by fitting the quadratic function to the maximum of the DCF. Error estimation of all values was done using the Monte Carlo method.

\section{Results}
 \begin{figure*}[h!]
 \centering
   \includegraphics[width=0.93\textwidth]{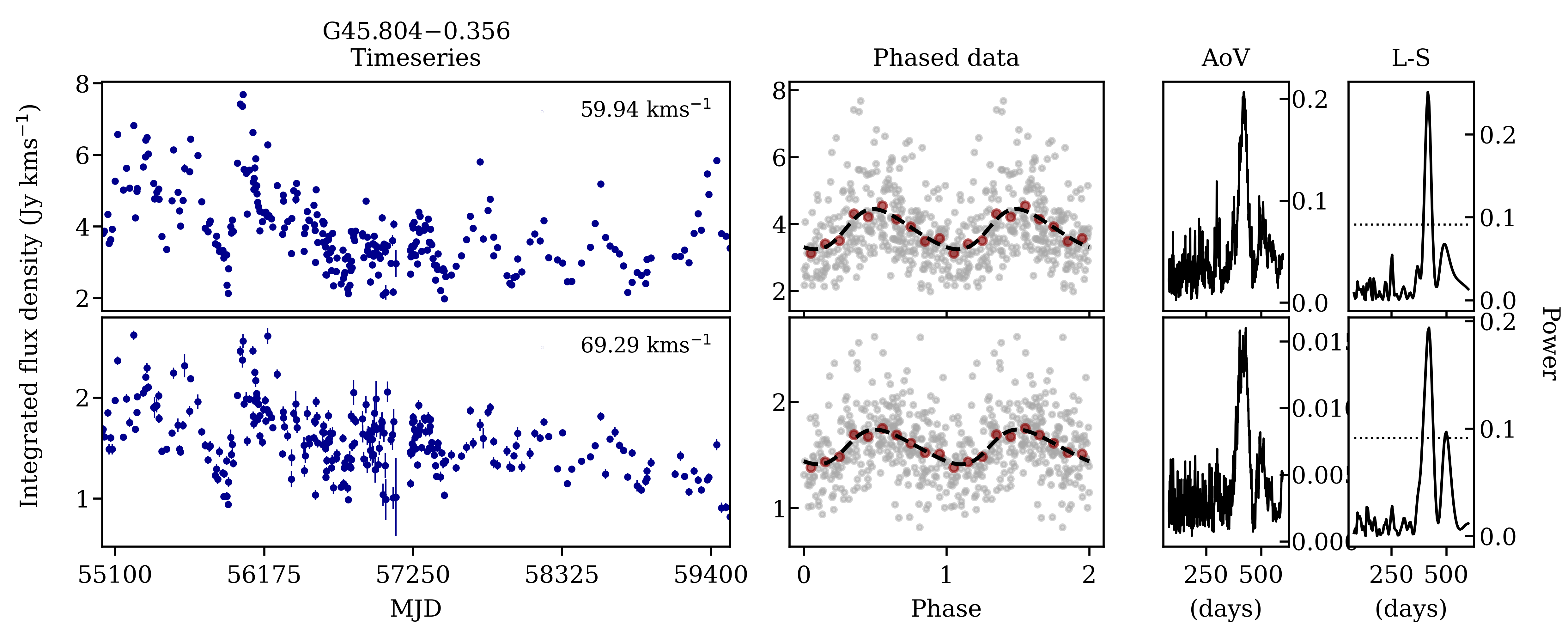}
    \includegraphics[width=0.93\textwidth]{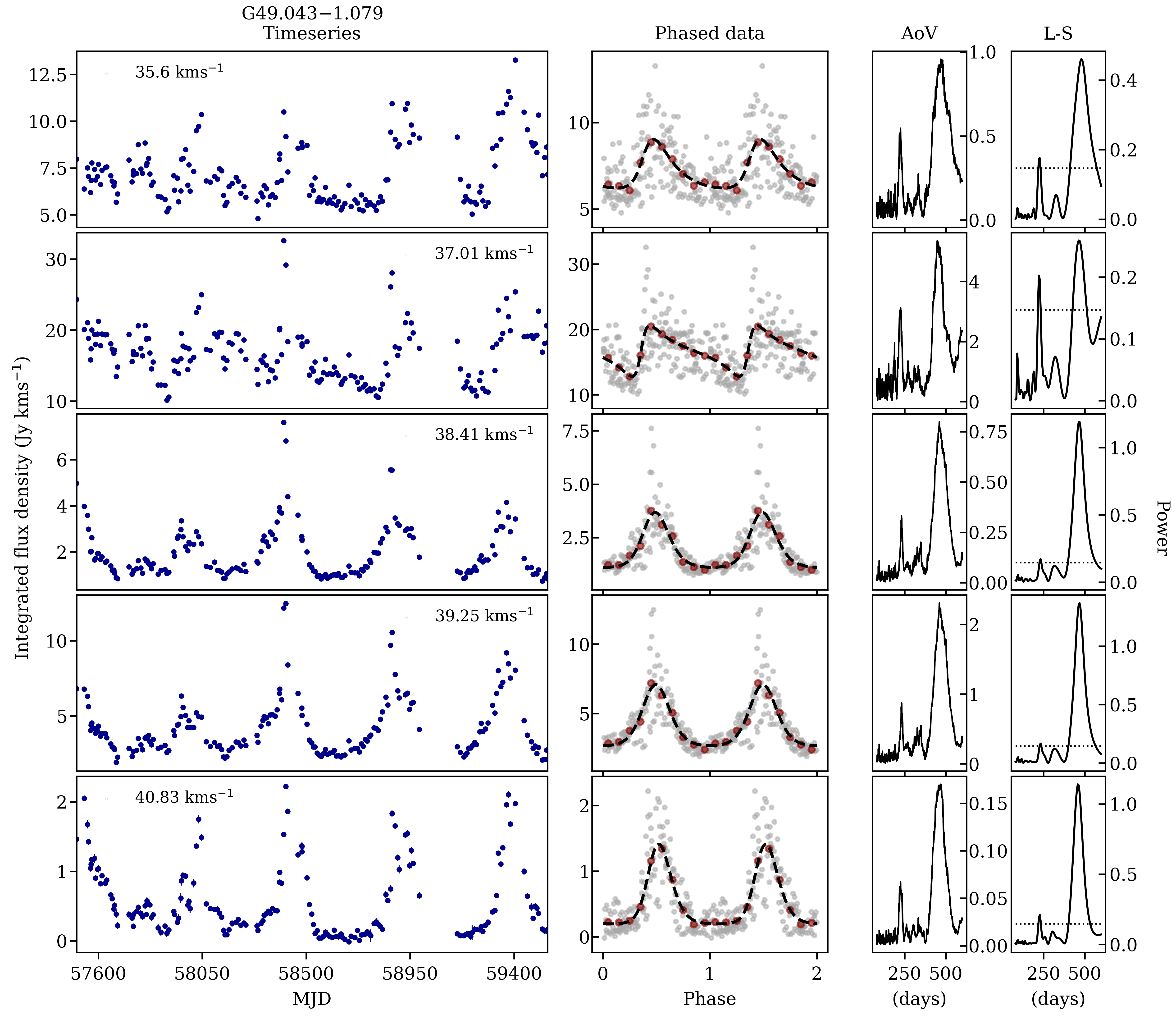}

   \caption{Periodic behaviour of both 6.7\,GHz sources. Panels in the leftmost column show the time series of spectral features. In the second column, grey points represent time series phased to the calculated period, red points show data binned with 0.1 phase intervals, and dashed black lines show the fitted asymmetric periodic power function. The third and fourth columns respectively show AoV and L-S periodograms. The dotted line in the last column represents the 1$\%$ false alarm probability level.}
   \label{fig:timeseries}
 \end{figure*}

\begin{table*}
  \caption{Parameters obtained from periodic analysis.}

    \label{Tab:line_properties}
    \centering
    \begin{tabular}{c l c c c l c l}
    \hline
     $V$ & $\overline{S}$ & R &  $P_{\mathrm{LS}}$ & P$_{\mathrm{AoV}}$ & FWHM & $R_\mathrm{rd}$ & $\tau$ \\
     (\kms) & (Jy) & & (days) & (days) & (days) & &  (days)\\
    \hline 
    \multicolumn{8}{l}{G45.804-0.356} \\
                  \hline
                   59.94 & 13.47 & 0.37 (0.07)& 416.76 (14.70) & 418.80 (23.38) & 199.00 (18.56) & 0.69 (0.19) & $-$\\
                   69.29 & 2.25 & 0.23 (0.04) & 416.95 (22.41) & 420.70 (32.62) & 208.28 (18.05) & 0.66 (0.22) & $-$\\
                  \hline
    \multicolumn{8}{l}{G49.043-1.079} \\
                  \hline
                   35.60 & 14.34 & 0.45 (0.06)& 480.64 (48.70) & 474.30 (53.45) & 148.20 (17.81) & 0.53 (0.13) & 67.7 (1.1)\\
                   37.01 & 14.08 & 0.61 (0.10) & 467.96 (39.23) & 459.41 (36.50) & 233.76 (21.47) & 0.20 (0.05) & 16.7 (15.0)\\
                   38.41 & 4.42 & 2.31 (0.15) & 468.18 (36.07) & 468.72 (41.84) & 133.78 (7.88)  & 0.87 (0.13) & 0\\
                   39.25 & 6.26 & 1.63 (0.13)& 468.53 (31.67) & 470.45 (42.14) & 137.82 (7.73)  & 0.97 (0.13) & 2.8 (1.1)\\
                   40.83 & 1.14 & 6.16 (0.38) & 461.02 (30.80) & 461.37 (37.59)& 115.75 (6.58)  & 0.81 (0.13) & 14.0 (1.7)\\
    \hline
    \end{tabular}
    \tablefoot{$V$ stands for the velocity of the spectral feature, $\overline{S}$ the average flux density during the observation period, R the relative flare amplitude calculated as $(S_{max}-S_{min})/(S_{min})$, $P_{LS}$ the period estimation using an L-S periodogram, P$_{AoV}$ the period estimation using AoV, FWHM the timescale of the flare,  $R_\mathrm{rd}$ the rise-to-decay ratio of the flare, and $\tau$ the time delay calculated with the DCF.}
\end{table*}

Results of the analysis are summarized in Fig.~\ref{fig:timeseries} and Table~\ref{Tab:line_properties}.\\

\begin{figure}
   \centering
   \includegraphics[width=\columnwidth]{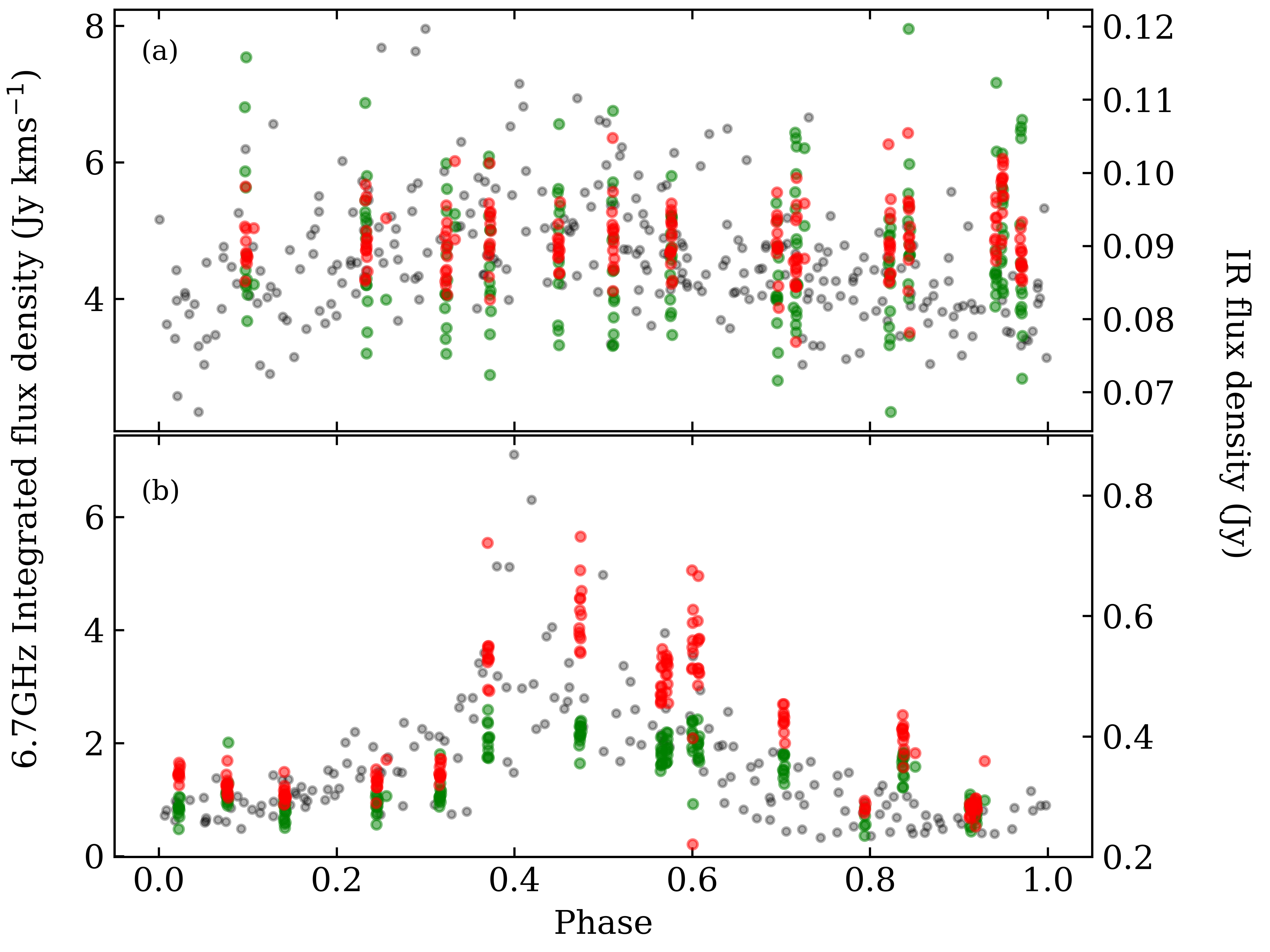}
   
   \caption{Comparison between the phased 6.7\,GHz line brightness and NEOWISE photometric data. Grey points show the methanol line for one of the chosen spectral features, green points represent 3.4 $\mu$m (W1),  and red points 4.6 $\mu$m (W2). The W1 flux was scaled up by a factor of 6. Panel (a) shows the G45.804$-$0.356 flux at 59.94 \kms, and panel (b) the source G49.043$-$1.0.79 feature 38.41\kms. }
   \label{fig:ir}
 \end{figure}

\noindent
\textbf{G45.804$-$0.356 }
The spectrum of this source consists of one strong feature at 59.94\,\kms\  and several weak ones in a velocity range from 55 to 71\,\kms\,  (Fig.\ref{fig:g45_dynamic}). From those weak features, only one, at 69.29\,\kms\,, was strong enough to be included in the analysis. Both this and the strong feature show periodic flares with an average L-S period of 416.9\,days, an AoV period of 419.8\,days, and a FWHM of 203.6\, days. The flare profiles are asymmetric, with rise-to-decay time ratios, $R_{\mathrm{rd}}$, of 0.69 and 0.66, respectively, a small relative amplitude of < 0.37, and no obvious quiescence period. No time lag measurements were carried out because the feature at 69.26\,\kms\,  is faint and flares with a small amplitude. During the monitoring period, no visible changes in the spectral distribution or flare amplitude were visible. We found no significant variations in flux in the IR data (Fig.\ref{fig:ir}).\\

\noindent
\textbf{G49.043$-$1.079 }
The entire spectral structure of the source with emission in velocities from 35 to 41\,\kms\, shows periodic variations (Fig. \ref{fig:g49_dynamic}).  Due to the complex structure and velocity drifts of  the blended features in velocities between 36 and 38\,\kms, the entire range was integrated as a single feature and omitted in further statistics. The periodic analysis resulted in an average L-S period of 469.3 days and an AoV period of 466.9 days. Features at 35.60\,\kms\, and 37.01 \kms\, show  asymmetric flare profiles, with $R_\mathrm{rd}$  of 0.53 and 0.20. The remaining features have symmetric flares within the margin of errors. The average length of the FWHM of the flares is 153.7\,days, and  relative amplitudes are as high as 6.16 in the 40.83\,\kms\ feature. Measured time lags with respect to the 38.41\,\kms\, feature range from 2.8 to 67.7\,days. A visual inspection of the time series revealed variations in the shape and amplitudes between the individual flares. There is a visible quiescence period between flares in all features except that at 37.01\,\kms. A phased light curve from NEOWISE data (Fig.\ref{fig:ir}) shows an IR flare simultaneous to the 6.7\,GHz variability. 

\section{Discussion}

Most of the 26 previously known periodic maser sources have periods that vary between 100 and 300\,days. Only four objects showed periods longer than 400 days (\citealt{Olech2019}, Table A3). The two sources presented in this paper,  with periods 416.9 and 469.3\,days, extend this small group and increase the total number to 28 sources.

In both cases, the quality of the flare profile analysis and the precision of the period estimation are limited by a relatively short observation period. For G45.804$-$0.356, the limiting factors are low flare amplitude and a high scatter of data before $\sim$57300 MJD. In G49.043$-$1.079, the observation period covers only four obvious flares, with the first being much weaker than the last three. On the other hand, the flare asymmetry of the 35.6\kms\, feature could  result from less pronounced variability at the beginning of the observation period and the influence of blended emission. The 37.01\,\kms\, feature shows an asymmetrical profile with flares that last much longer. Strong velocity drifts visible during the monitoring suggest that multiple maser cloudlets make up this emission. The resulting cumulative flare profile could be significantly different if the cloudlets flared with a slight time lag due to the line of sight effect. This possibility, combined with a small flare amplitude at the beginning of observations, is the most likely cause of this unusual profile. Continued monitoring will help better characterize the behaviour of both sources.
\subsection{Source morphology}
The 6.7\,GHz emission of both sources was imaged by \cite{Hu2016} with the Very Large Array (VLA)  in C configuration (maps from their work are presented in Fig. \ref{fig:hu}). Although due to a short on-source time of 20s and low resolution (beam size of $3"\times6"$) the position of individual maser cloudlets has a high error, these maps can provide us insights into the nature of these sources. The map of G45.804$-$0.356 includes only a part of the emission seen in single-dish spectra, with only the 59.94\kms\, feature visible. A weak blue-shifted emission seen at a distance of $\sim1.3"$ (i.e $>$9000\,au at a distance of 7.3$^{+1.5}_{-1.0}$ kpc; \citealt{Reid2019}) most likely belongs to a different source. With this incomplete structure, it is impossible to analyse its morphology. Additionally, no continuum emission was detected with a limit of 45 $\mu$Jy beam$^{-1}$.
For source G49.043$-$1.079, all single-dish features were mapped, and a linear structure with a velocity gradient is visible. With a parallax distance of 6.1$^{+0.9}_{-0.7}$ 
%(+0.94/-0.73) 
kpc \citep{Reid2019} and an angular size of $<0.3"$, the emission has a linear size of $\sim$ 1800\,au, which is consistent with typical 6.7\,GHz maser source sizes. The structure of emission hints at edge-on disc morphology. A weak continuum emission of 0.23\,mJy was detected in this source.
\subsection{Periodicity causes} 
A number of hypotheses have been proposed to explain periodic behaviour in masers. Here we discuss them in the context of the presented sources.

The CWB model was proposed and successfully applied to a few sources by \citet{vanderWalt2011} and \citet{vanderWalt2016}. It considers a periodic wind interaction in binary systems that generates additional ionizing radiation waves. Consequently, this leads to a change in seed photon flux and, in turn, observed 6.7\,GHz brightness. This model produces an asymmetric flare profile with a rapid brightness increase and a longer decrease phase. Nevertheless, it requires the existence of an ultra-compact HII region. In the case of both sources presented in the paper, CWB is not the likely cause. G45.804$-$0.356 shows no continuum emission typical to HII regions. Although G49.043$-$1.079 has weak continuum emission, the flare profiles show mostly symmetrical Gaussian profiles, which this model does not readily produce.

Numerical simulations of massive protostars undergoing substantial accretion have revealed that they could become pulsationally unstable \citep{Inayoshi2013}. The star's luminosity varies periodically, increasing the local dust temperature and leading to a rise in the pumping efficiency of the maser and its brightness. The relation between the pulsation period and the protostellar luminosity is given by the relation log(L) = 4.62 + 0.98 log(P /100 days), where L is luminosity in L$_{\odot}$ and P is the period. Since G45.804$-$0.356 shows no detectable variability in IR, we can rule out this process. In the case of G49.043$-$1.079, observed IR flares might suggest periodic pulsations. For {a} period of 469 days, the predicted luminosity should be $\sim 2\times10^{5} L_{\odot}$, but the luminosity estimated from observations is $\sim 2\times10^{3} L_{\odot}$  \citep{urquhart2018}.  The difference of two orders of magnitude suggests that a periodic pulsation scenario is unlikely.

According to the rotating spiral shock  model proposed by \cite{Parfenov2014}, matter accreting onto the binary system does so through hot, shocked, spiral structures, which are seen in many simulations \citep{Artymowicz1996,Munoz2016}. As an effect of the rotation, these shocks would periodically heat up portions of the accretion disc, increasing the pumping efficiency. The geometry of the system strongly constrains this model as it requires an edge-on configuration. Additionally, the model requires neither the presence of detectable continuum emission nor strong variability in IR brightness. It appears as the most probable cause for G45.804$-$0.356. However, the lack of a complete, high angular resolution map of maser emission prevents us from verifying this hypothesis. In the case of G49.043$-$1.079, the edge-on morphology could allow this mechanism, but it is unclear if the spiral shocks could produce an observed IR flare.

Interaction between a binary system and accreting matter does not only lead to the creation of spiral structures in the circumbinary disc: in some cases, it can also strongly modulate accretion \citep{Munoz2016}. We have argued that this process is responsible for the periodic variability of G107.298$+$5.639 and G59.633$-$0.192 \citep{szymczak2016,Olech2019}. Well-separated features of G49.043$-$1.079 show methanol and IR flare profiles very similar to those of G59.633$-$0.192 mentioned above; the two sources most likely share the same driving mechanism.

A recent study of three periodic masers, G9.62+0.19E, G22.357$+$0.066, and G25.411$+$0.105, suggests that source geometry and orientation in the plane of the sky might also be an important factor influencing observed flare profiles \citep{morgan2021}, but more research is needed. High-quality data limited to single-dish observations are presented for the sources analysed in this paper. However, interferometric spectral and continuum observations with milliarcsecond resolution are needed to fully understand these objects.

\subsection{Other characteristics}
The line of sight effect cannot easily explain the large time lags between features of G49.043$-$1.079. A time lag of 67.7 days would require a distance of 12000\,au, which is an order of magnitude higher than the linear size of the emission in the sky plane and a typical size for methanol maser sources. A similar phenomenon was observed in three other periodic sources, G25.411$+$0.105 \citep{Szymczak2015}, G30.400$-$0.296 \citep{Olech2019}, and G331.13$-$0.24 \citep{Goedhart2014}, and is not yet understood. Large inhomogeneities in the local environment and differences in dust and gas heating times might be the causes. Other noteworthy phenomena visible in the source are rapid jumps in the intensity (shorter than two weeks), which are most pronounced during flare maxima at 38.41 and 39.25\,\kms. A similar phenomenon was first detected in the quasi-periodic source G33.641$-$0.228 \citep{fujisawa2014}, but more frequent observations are needed to verify its nature.

\section{Summary}
We have presented the analysis of two previously unidentified periodic 6.7\,GHz methanol masers, G45.804$-$0.356 and G49.043$-$1.079, which respectively have periods of 416.9 and 469.3 days. Infrared variability in source G49.043$-$1.079 shows a periodic flare that is simultaneous with methanol maser changes. Measured phase lags between features in this source are as high as 68 days, which cannot be attributed to line of sight effects.
No clear explanation can be inferred for the cyclical variability in G45.804$-$0.356 with the available data. For G49.043$-$1.079, periodicity is most likely caused by a modulated accretion leading to an increased pumping efficiency of methanol masers. Both sources should be targeted in future Very Long Baseline Interferometry observations in order to examine and verify our conclusions. 

\begin{acknowledgements}
M.O. thanks the Ministry of Education and Science of the Republic of Poland for support and granting funds for the Polish contribution to the International LOFAR Telescope (arrangement no 2021/WK/02) and for maintenance of the LOFAR PL-612 Baldy (MSHE decision no. 59/E-383/SPUB/SP/2019.1)\\
The 32\,m radio telescope is operated by the Institute Astronomy, Nicolaus Copernicus University and supported by the Polish Ministry of Science and Higher Education SpUB grant. We thank the staff and students for assistance with the observations.
This research made use of Astropy,\footnote{http://www.astropy.org} a community-developed core Python package for Astronomy.
This publication makes use of data products from the Near-Earth Object Wide-field Infrared Survey Explorer (NEOWISE), which is a joint project of the Jet Propulsion Laboratory/California Institute of Technology and the University of Arizona. NEOWISE is funded by the National Aeronautics and Space Administration. This research has made use of NASA’s Astrophysics Data System Bibliographic Services.
\end{acknowledgements}

\bibliographystyle{aa}
\bibliography{biblio}

\begin{appendix}
\section{Supplementary figures}
\begin{figure*}[ht]%
    \centering
    \subfloat{{\includegraphics[width=0.45\textwidth]{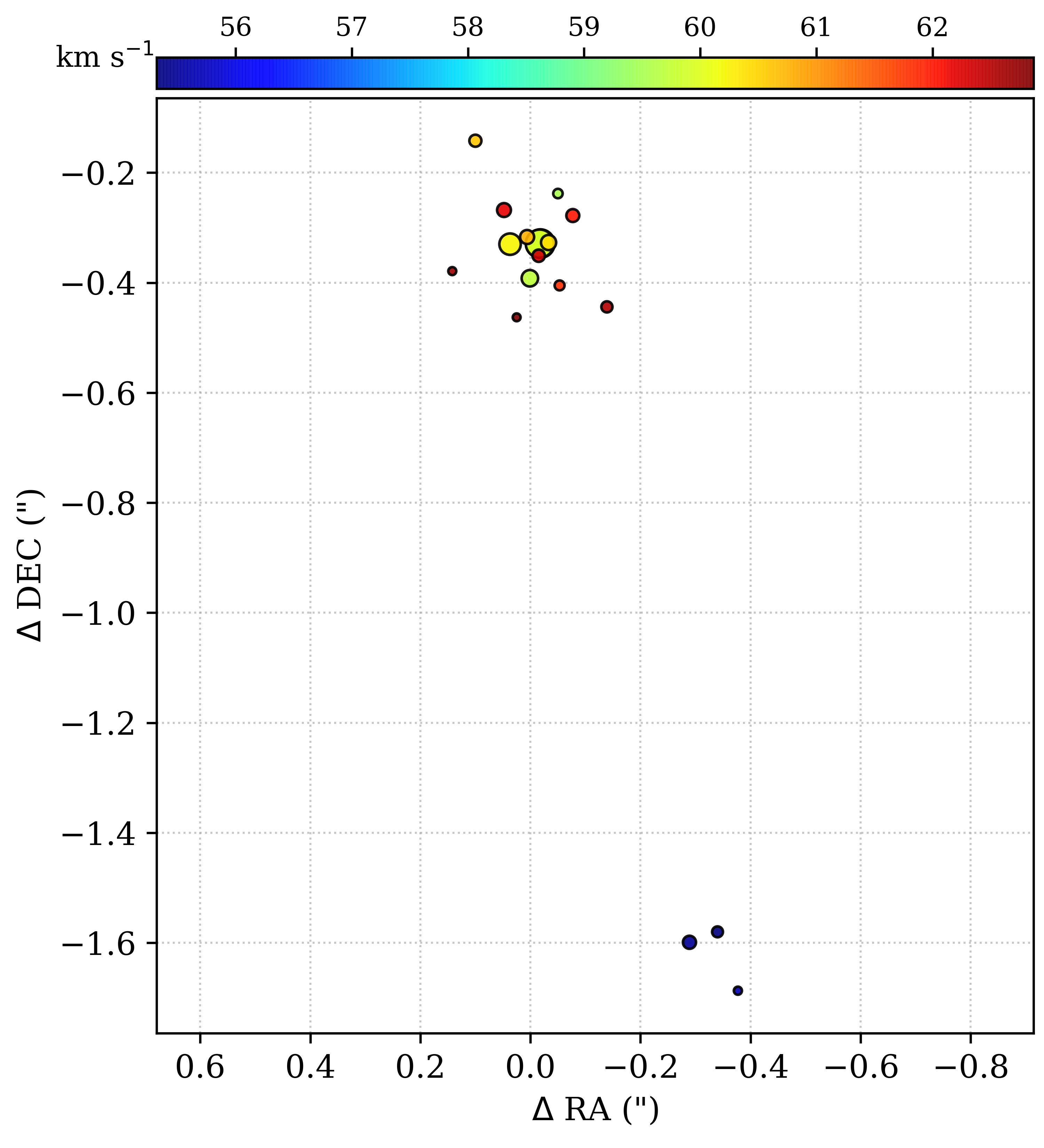} }}%
    \qquad
    \subfloat{{\includegraphics[width=0.45\textwidth]{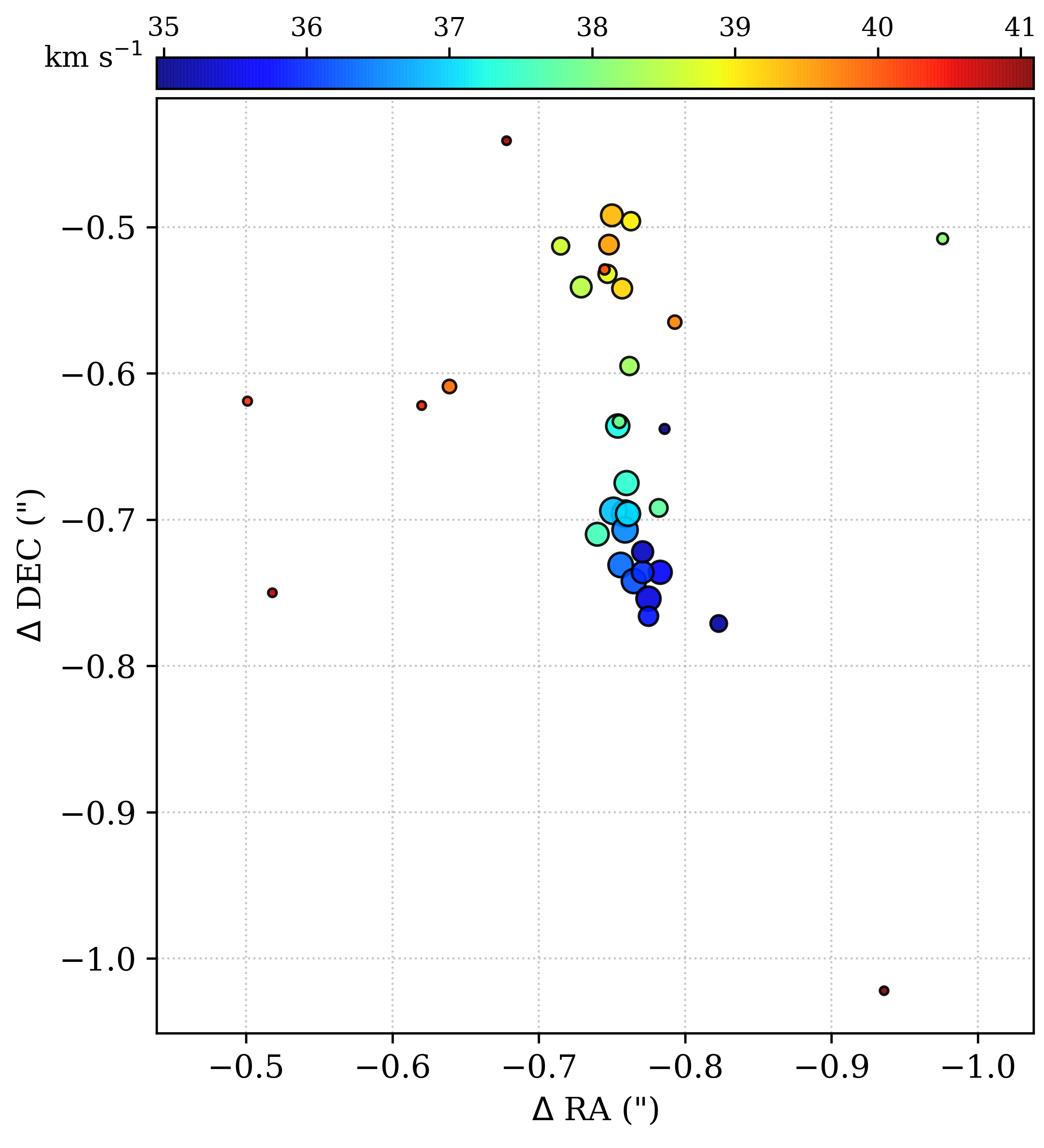} }}%
    \caption{VLA map of 6.7\,GHz methanol maser emission in the sources G45.804$-$0.356 (left) and G49.043$-$1.079 (right) taken from \cite{Hu2016}. The size of the circles is scaled logarithmically with measured brightness, and their colours represent velocity.}%
    \label{fig:hu}%
\end{figure*}

\end{appendix}

\end{document}